\documentclass[pdflatex,sn-mathphys-num]{sn-jnl}% Math and Physical Sciences Numbered Reference 
\usepackage{indentfirst,csquotes}

\usepackage{graphicx}%
\usepackage{multirow}%
\usepackage{amsmath,amssymb,amsfonts}%
\usepackage{amsthm}%
\usepackage{mathrsfs}%
\usepackage[title]{appendix}%
\usepackage{xcolor}%
\usepackage{textcomp}%
\usepackage{manyfoot}%
\usepackage{booktabs}%
\usepackage{algorithm}%
\usepackage{algorithmicx}%
\usepackage{algpseudocode}%
\usepackage{listings}%
\usepackage{graphicx}% Include figure files
\usepackage{amsmath,amssymb,bm, bbold}% bold math
\usepackage{xr}
\usepackage{color}
\usepackage[normalem]{ulem}
\usepackage{wasysym}
\usepackage{siunitx}
\usepackage{physics}
\usepackage{lineno}
\usepackage{gensymb}
\usepackage[T1]{fontenc}
\usepackage{booktabs}
\usepackage[section]{placeins}
\usepackage{textcomp}
\usepackage{comment}
\usepackage{upgreek}
\usepackage{textgreek}
\usepackage{romannum}
\AtBeginDocument{\RenewCommandCopy\qty\SI}

\theoremstyle{thmstyleone}%
\theoremstyle{thmstyletwo}%

\theoremstyle{thmstylethree}%

\newcommand{\moire}{moir\'e }
\newcommand{\Moire}{Moir\'e }
\newcommand{\MoSe}{MoSe$_2$ }

\begin{document}

\title[Article Title]{\textbf{\Moire exciton polaron engineering via twisted hBN}}

\author[1,2]{\fnm{Minhyun} \sur{Cho}}
\author[2]{\fnm{Biswajit} \sur{Datta}}
\author[1]{\fnm{Kwanghee} \sur{Han}}
\author[3]{\fnm{Saroj B.} \sur{Chand}}
\author[2]{\fnm{Pratap Chandra} \sur{Adak}}
\author[2]{\fnm{Sichao} \sur{Yu}}
\author[4]{\fnm{Fengping} \sur{Li}}
\author[5]{\fnm{Kenji} \sur{Watanabe}}
\author[6]{\fnm{Takashi} \sur{Taniguchi}}
\author[7]{\fnm{James} \sur{Hone}}
\author[4]{\fnm{Jeil} \sur{Jung}}
\author[3,8]{\fnm{Gabriele} \sur{Grosso}}
\author*[1,9]{\fnm{Young Duck} \sur{Kim}}\email{ydk@khu.ac.kr}
\author*[2]{\fnm{Vinod M.} \sur{Menon}}\email{vmenon@ccny.cuny.edu}

\affil[1]{\orgdiv{Department of Physics}, \orgname{Kyung Hee University}, \orgaddress{\street{26 Kyungheedae-ro}, \city{Seoul}, \postcode{02447}, \country{ Republic of Korea}}}

\affil[2]{\orgdiv{Department of Physics}, \orgname{City College of New York}, \orgaddress{\city{New York}, \postcode{10031}, \state{NY}, \country{United States}}}

\affil[3]{\orgdiv{Photonics Initiative}, \orgname{Advanced Science Research Center, City University of New York}, \orgaddress{\city{New York}, \postcode{10031}, \state{NY}, \country{United States}}}

\affil[4]{\orgdiv{Department of Physics and Department of Smart Cities}, \orgname{University of Seoul}, \orgaddress{\city{Seoul}, \postcode{02504}, \country{Republic of Korea}}}

\affil[5]{\orgdiv{Research Center for Electronic and Optical Materials}, \orgname{National Institute for Materials Science}, \orgaddress{\street{1-1 Namiki}, \city{Tsukuba}, \postcode{305-0044}, \country{Japan}}}

\affil[6]{\orgdiv{Research Center for Materials Nanoarchitectonics}, \orgname{National Institute for Materials Science}, \orgaddress{\street{1-1 Namiki}, \city{Tsukuba}, \postcode{305-0044}, \country{Japan}}}

\affil[7]{\orgdiv{Department of Mechanical Engineering}, \orgname{Columbia University}, \orgaddress{\city{New York}, \postcode{10027}, \state{NY}, \country{United States}}}

\affil[8]{\orgdiv{Physics Program, Graduate Center}, \orgname{City University of New York}, \orgaddress{\city{New York}, \postcode{10016}, \state{NY}, \country{United States}}}

\affil[9]{\orgdiv{Department of Information Display}, \orgname{Kyung Hee University}, \orgaddress{\city{Seoul}, \postcode{02447}, \country{Republic of Korea}}}

\abstract{\textbf{Twisted hexagonal boron nitride (thBN) exhibits emergent ferroelectricity due to the formation of moiré superlattices with alternating AB and BA domains. These domains possess electric dipoles, leading to a periodic electrostatic potential that can be imprinted onto other 2D materials placed in its proximity. Here we demonstrate the remote imprinting of moiré patterns from twisted hexagonal boron nitride (thBN) onto monolayer \MoSe and investigate the resulting changes in the exciton properties. We confirm the imprinting of moiré patterns on monolayer \MoSe via proximity using Kelvin probe force microscopy (KPFM) and hyperspectral photoluminescence (PL) mapping. By developing a technique to create large ferroelectric domain sizes ranging from 1 µm to 8.7 µm, we achieve unprecedented potential modulation of \(\sim\) 387±52 meV. We observe the formation of exciton polarons due to charge redistribution caused by the antiferroelectric moiré domains and investigate the optical property changes induced by the moiré pattern in monolayer \MoSe by varying the moiré pattern size down to \(\sim\)110 nm. Our findings highlight the potential of twisted hBN as a platform for controlling the optical and electronic properties of 2D materials for optoelectronic and valleytronic applications. }}

\maketitle

\section{Introduction}
    van der Waals (vdW) ferroelectrics with two-dimensional (2D) crystal lattices has emerged as a promising platform for applications in next generation quantum electronic and spintronic devices\cite{wang2023towards}. In addition to naturally occurring layered vdW ferroelectrics, there exists the intriguing possibility of achieve ferroelectricity in vdW heterostructures made of parent compounds that are not intrinsically ferroelectric\cite{zhang2023ferroelectric}. Hexagonal boron nitride (hBN) is a good example of such emergent ferroelectricity that arises from the sliding or twisting between two hBN layers\cite{li2017binary}\cite{vizner2021interfacial}\cite{woods}.
    When two hBN layers are stacked with a slight misalignment, which is twisted hBN (thBN), the nitrogen and boron at the interface of hBN form a triangular moiré superlattice structure with AB and BA domains, exhibiting an electric dipole moment. The out of plane dipole moment has opposite directions in the AB and BA domains because of the different atomic arrangement, resulting in antiferroelectric \moire domains. Along with the ferroelectric properties, the possibility to tune the properties on demand via nanomechanical rotation\cite{yao2021enhanced} or via external electric field to induce sliding between AB and BA domains shows promise for a programmable ferroelectric platform\cite{yasuda2021stacking}\cite{yasuda2024ultrafast}.

    2D materials such as graphene, and transition metal dichalcogenides (TMDC) that host unique electronic and optical properties are also highly susceptible to external environment. Therefore, there has been significant research aimed at manipulating these properties by engineering the external environment\cite{raja2017coulomb}\cite{forsythe2018band}\cite{lee2021boosting}. 
    Recently, studies have been conducted on imprinting \moire patterns onto different target 2D materials to investigate the effect of remote interactions\cite{tp3}\cite{tp2}\cite{tp1}\cite{wang2024band}. There have also been recent reports on imprinting \moire patterns on target 2D materials placed in the proximity of thBN resulting in localized excitons and doping\cite{zhao2021universal}\cite{kim2024electrostatic}\cite{Doping3}. However, there has been no clear evidence of the imprinted \moire patterns and the modified optical properties induced by the antiferroelectric \moire domains on the proximitized excitons of the target material. 

    Here, we report on the optical properties of monolayer \MoSe with imprinted moiré patterns from the underlying thBN. \Moire patterns with small twist angles ranging from 0.01 to 0.2 degrees were created by stacking two nearly aligned hBN flakes. It was also found that twist angles below 0.01 degrees can be realized by repeating the stacking process on the thBN. Using this approach, we achieved unprecedented potential modulation of \(\sim\)387±52 meV. We confirmed the imprinting of \moire patterns on monolayer \MoSe via proximity, using Kelvin probe force microscopy (KPFM) and hyperspectral PL mapping to visualize and quantify the effect. Finally, we investigated the changes in exciton properties due to the charge redistribution caused by the antiferroelectric \moire domains resulting in the formation of exciton polarons. These are excitons dressed by the presence of additional charges of the Fermi sea\cite{fey2020theory}\cite{sidler2017fermi}\cite{smolenski2019interaction}\cite{goldstein2020ground}\cite{Doping1}\cite{huang2023quantum}. In TMDC, this interaction results in the splitting of the bare exciton into attractive polaron branch occupying the trion state in the Fermi sea, and the repulsive polaron branch corresponding to the bare exciton state in the Fermi sea. We confirmed the Fermi level change in monolayer \MoSe induced by the \moire potential of the thBN substrate from the energy shift of exciton polarons. 
   %Due to the antiferroelectric \moire-domains of thBN, exciton-polarons are formed in the \moire domains showing large blue shift of 8.55 meV for the repulsive polaron branch and .

\section{Results and Discussion}
    Schematic of the sample structure is shown in Figure 1(a), where \moire patterns form at the interface of the two hBN layers. When the arrangement of atoms is in a parallel configuration it results in three atomic registries : AA, AB and BA. In this case, AB and BA configurations relax to the most stable state, forming triangular antiferroelectric \moire domains. The AB domain has a dipole moment in the downward direction, while the BA domain has a dipole moment in the upward direction. The different directions of the dipole moments induce an electric field on the surface. This in turn leads to electron redistribution in the target 2D material,  MoSe$_2$, as shown schematically in Figure 1(a). These electrons interact with the excitons to form exciton polarons. In a previous work, it was shown that the target material placed on thBN can be doped not only by the potential difference but also by the in-plane induced electric field\cite{Doping3}. The schematic of this in-plane electric field is illustrated in Figure 1(b). Additionally, the direction of the in-plane electric field applied between the AB and BA domains can vary depending on whether the target material is positioned above or below the thBN. When \MoSe is positioned above, electrons migrate from the AB domain to the BA domain. As a result, in Figure 1(c), the Fermi energy increases only in the BA domain, and in this specific domain, excitons interact with electrons to form exciton polarons while the AB domain contains only bare excitons. When the \MoSe monolayer is positioned below the thBN, the direction of the electric field is reversed, resulting in the formation of the exciton polaron in the AB domain and bare excitons in the BA domain. The difference in Fermi levels between the domains lead to different interactions between excitons and electrons, resulting in an energy difference depending on the domain. In MoSe$_2$, there are excitonic and trionic states which undergo an energy shift due to the repulsive and attractive interactions with electrons, respectively.
    The energy difference between the repulsive polaron and the attractive polaron can be approximated as the sum of the trion binding energy and the Fermi energy as shown in Figure 1(d)\cite{huang2023quantum}. To confirm the formation of a \moire pattern on monolayer MoSe$_2$, we performed KPFM measurements on the proximitized \MoSe surface. 
    
    The fabrication of thBN involves stacking two hBN layers that are exfoliated in close location and aligned nearly perfectly in their as-exfoliated state\cite{woods} using the PC (Polycarbonate) dry transfer method\cite{zomer2014fast}. The \moire size made by this method is usually 100 nm to 1µm, corresponding to twist angles varying from 0.0143 to 0.143 degrees, resulting in 100 to 250 meV potential difference (Figure S1). The bulk \MoSe was exfoliated onto a PDMS substrate and the monolayer \MoSe is transferred directly from the PDMS substrate to thBN. The substrate is grounded for KPFM measurements with positive bias through the tip. Figures 1(e) and 1(f) show the KPFM image of thBN and the monolayer \MoSe on the thBN, respectively. It exhibits \moire sizes about 161±30 nm with the monolayer \MoSe showing higher noise because of the charge redistribution in the MoSe$_2$. Figure 1(g) shows line cuts of the KPFM image for the two scenarios indicating excellent agreement both in the induced potential difference around 110 meV as well as the \moire size. This indicates that the \moire patterns did not change during the transfer process and that the \moire potential is imprinted on the proximitized MoSe$_2$. Subsequently, Raman spectroscopy was conducted to investigate any strain effects arising from the underlying \moire structure (Figure S4). Both the peak energy of $A_{1g}$ and $E_{2g}^1$ modes showed no change. This confirmed that no additional strain was introduced to \MoSe by the underlying \moire superlattice.

%[NEED to motivate the reader for why we need larger domain size: - that is for spectroscopy]

    The effects on the exciton properties of such small domain size (161±30 nm) cannot be appreciated with standard spectroscopic methods as the effects are average over several domains (Figure S4). To realize domains that can be investigated using far field measurements, we developed a technique to realize large ferroelectric domain sizes ranging from 1 µm to 8.7 µm. This technique relies on the tendency of the thBN to consistently relax to the most stable state of 0 degrees when the stacking process is repeated. Figure 2(a) shows the KPFM image of the sample with the largest \moire size among the stacked devices. To check for additional relaxation through heating, we annealed the sample and then conducted KPFM measurements (Figure S2). The \moire size remains unchanged after annealing. When the second transfer process was conducted on the same substrate, as shown in Figure 2(b), the \moire size increased to a maximum of 8.7 µm. The angle moved closer to 0 degrees only after we repeat the transfer. 
    
    This relaxation to 0 degrees is consistent with studies of bulk BN microrotators, where 0 degree and 60 degree configurations are the most stable\cite{yao2021enhanced}. In this range of aligned angles, rigid \moire patterns relaxed into specific shapes. PFM studies of these patterns reveal strain accumulation at domain boundaries\cite{woods}\cite{chiodini2024electromechanical}. In the second stacking process, the two hBN flakes relax into large AB and BA domains with minimal domain boundaries, thus reaching their most stable configuration. In Figure 2(c), we compare potential difference before and after the second transfer process. Following the second transfer, potential difference exhibits a much larger value of 387±52 meV compared to the initially formed potential difference of 242±30 meV. 
   
    As the \moire size increases, the Coulomb interaction between \moire domains decreases. Consequently, the observed \moire potential is much larger than previous reports\cite{zhao2021universal}. In Figure 2(d), we show the maximum potential difference for the three devices after the first and second transfers. After the second transfer, the size increases, resulting in a range of \moire sizes, with the maximum moiré size indicated. The potential before and after transfer in Figure 2(c) corresponds to device 3. Similar KPFM results for other devices are included in Figure S2. The magnitude of the \moire potential is proportional to the strength of the dipole moment in the thBN, according to the following equation\cite{zhao2021universal}:
   \[V_{max}=\frac{P_{max}}{\epsilon_0 }e^{-Gz}\]
   Where, $V_{max}$ is the maximum potential difference between AB and BA domains, $P_{max}$ is the maximum interlayer charge polarization, $\epsilon_0$ is the dielectric constant, $G$ is $4\pi /\sqrt{3}a$, where $a$ is \moire period size, and z is distance from target material to interfaces of two hBN. Previously, the maximum potential difference observed was \(\sim\) 203 meV when $P_{max}$ value is around 2 pC/m. This value agrees well when the \moire size is below 1 µm. In contrast, following the second transfer, the \moire size increases and the maximum potential difference we observe is \(\sim\) 387±52 meV, which is 1.91 times larger than domains formed after single transfer.

    The realization of larger \moire domains allowed us to carry out hyperspectral imaging of the exciton emission (see Methods). Figure 3(a) shows the results of these measurements where thBN underwent a second transfer onto MoSe$_2$/hBN to create large \moire domains allowing us to spectroscopically investigate the exciton properties in the different \moire domains. We observed distinct PL emisison pattern from the AB and BA domains. In the Figure 3(b), we show the intensity ratio where the trion state intensity is normalized to the exciton state intensity. Furthermore, the hyperspectral imaging clearly demonstrates the \moire patterns from the intensity, energy, and linewidth of the exciton and trion states in the far field (Figure S3). Figure 3(c) show the contour plot of PL spectra along the white arrow in Figure 3(b). The line cuts at the AB and BA domains as well as the domain boundary are shown in Figure 3(d) and the peak shift of two states ($A_{1S}^-$, $A_{1S}$) are shown in Figure 3(e). It clearly shows the peak shift along the domains with opposite direction. Exciton states ($A_{1S}$) shows blue shift about 6 meV and trion states ($A_{1S}^-$) shows red shift about 1 meV. The difference in Fermi levels between the domains results in distinct interactions between excitons and electrons, leading to domain-dependent energy shifts. In MoSe$_2$, both exciton and trion states exist, and their energy difference arises from repulsive and attractive interactions with electrons, respectively. The energy difference between the two states observed in the spectra of each domain in Figure 3(d) is shown in Figure 3(f). Based on this energy difference, we determined the Fermi level in each domain using the following equation\cite{huang2023quantum}: 
    $$\Delta E_{RP-AP} = E_{tB} + \frac{3}{2}{\epsilon_F}$$

    Where $E_{tB}$ is the trion binding energy and $\Delta E_{RP-AP}$ is the energey difference between the repulsive polaron (RP) and attractive polaron (AP). The Fermi level, $\epsilon_F$, increases along the domain with maximum increase in Fermi level estimated to be ~ 4.54 meV across the boundary. The trion binding energy, $E_{tB}$ is assumed to be a constant (20.85 meV) within this range of Fermi level shift and is obtained from the fit to the experimentally observed shift in 
    $\Delta E_{RP-AP}$. Here, we highlight that the peak shift is mostly due to the repulsive polaron. When \MoSe is positioned below thBN, the direction of the in-plane electric dipole points from the AB to the BA domain. The increased Fermi energy in the AB domain induces the formation of exciton polarons exclusively in this domain. Conversely, the excitons in the BA domain, where the Fermi energy is reduced, approaches the bare exciton state.
    
    In addition to the 1s excitons, the excited states of excitons (Rydberg) are also affected by the underlying \moire superlattice. Figure 4(a) shows a hyperspectral PL image of the $A_{2S}$ Rydberg exciton intensity, revealing distinct spatial distribution across the AB and BA domains. The photon energy along the white dotted arrow indicated Figure 4(a) is plotted as a contour map in Figure 4 (b) along with line cuts in the two domains showing the PL spectra in Figure 4(c). Similar to the $A_{1S}$ exciton, the $A_{2S}$ exciton also shifts in energy from 1.790 eV to 1.799 eV. Figure 4(d) shows that the peak shift of the Rydberg exciton polaron ($A_{2S}$) which is observed to be larger than that of the ground state exciton polaron ($A_{1S}$). This observation indicates that the higher energy Rydberg states are more sensitive to the \moire potential and exhibit stronger exciton-electron interaction compared to the ground state. This behavior is consistent with previous studies on Rydberg exciton polarons\cite{tp3}\cite{chernikov2014exciton}.

    Finally, we investigated the role of the \moire lattice size on the excitons in monolayer \MoSe especially in smaller domain sizes (below 1 µm). KPFM images in Figure 5(a) show four distinct moiré sizes on thBN, ranging from 111 nm to 252 nm. The corresponding PL spectra for both the repulsive polaron ($A_{1S}$) and the attractive polaron ($A_{1S}^-$) are shown in Figure 5(b) and 5(c), respectively. Due to the variation in trion binding energy and initial electron density among different samples, we normalized the energy difference from the moiré-less region for each sample. Additionally, the Fermi level change was estimated from the energy shift of the repulsive polaron, as previously shown in Figures 3(e) and (f). Figure 5(d) quantifies this relationship, plotting the PL peak shift as a function of the inverse moiré size ($1/\lambda$). Notably, the repulsive polaron ($A_{1S}$) exhibits a consistent blue shift, while the the attractive polaron ($A_{1S}^-$) displays a red shift, agreeing with previous observations\cite{Doping1}\cite{huang2023quantum}. In Figure 5(e), we determined the strength of the electric field as a function of \moire size using the maximum potential difference based on the \moire sizes in Figure 2(d). This calculation was performed using the following equation $E_{max} = (2\pi/\lambda)V_{max}$. As the \moire size decreases, the strength of the applied electric field increases sharply, supporting the behavior we observe. The inset in Figure 5(e) schematically depicts this charge redistribution, highlighting the accumulation of electrons in specific domains due to the enhanced electric field for smaller \moire sizes.
    
    In addition to the shift in exciton energies in the proximity of \moire superlattice, another intriguing effect is the modification in the exciton-exciton interaction due to the \moire potential\cite{zhang2021van}\cite{fitzgerald2022twist}. Pump power dependent PL measurements show changes in he saturation behavior of the PL of the excitons placed in the proximity of the \moire superlattice (Figure S4). The emission was found to saturate at lower pump power for the \moire proximitized excitons as compared to the bare excitons. Interestingly, the trion state did not show the saturation behaviors. Further investigation of exciton-exciton interaction modified by the underlying \moire potential will be the subject of a future work. A more detailed analysis using Rydberg states, which can probe exciton behavior more sensitively as shown in Figure 4, or further studies with gate-controlled Fermi levels are required in the future.

\section{Conclusion}
    In conclusion, we have demonstrated the imprinting of \moire patterns from twisted hBN onto monolayer \MoSe through KPFM and PL measurements. We have proposed a method for creating small \moire angles below 0.01 degrees by repeating the transfer process, allowing the realization of large domains and observe the excitons and exciton polaron behavior for the different domains through far-field spectroscopic measurements. The difference in Fermi levels between the domains result in distinct interactions between excitons and electrons, leading to domain-dependent energy shifts. The increased Fermi energy in the AB domain induces the formation of exciton polarons exclusively in this domain. Conversely, the excitons in the BA domain, where the Fermi energy is reduced, approaches the bare exciton state. Subsequently, we experimentally confirmed the optical property changes induced by the \moire pattern in monolayer \MoSe by varying the \moire pattern size down to 111 nm, which is attributed to the charge redistribution by the in-plane electric field. These results highlight the potential of twisted hBN as a platform for controlling the optical and electronic properties of 2D materials for optoelectronic and valleytronic applications.

% Experimental section
\section{Methods}
\medskip
\textbf{Sample preparation}
   thBN was fabricated by aligning the zigzag or armchair edge formed when the two hBNs are naturally exfoliated. hBN was exfoliated on a 90nm \textrm{SiO$_2$}/Si substrate using PC (Polycarbonate) dry transfer method. We exfoliated the flux-grown high quality \MoSe bulk crystal onto the PDMS and transferred to hBN or thBN. After transfer process, samples were annealed at 300 °C for 1 hour under a nitrogen environment and contact cleaned\cite{rosenberger2018nano} using AFM to remove strained regions, bubbles and the polymer residue. \\

\textbf{KPFM measurement}
   We performed KPFM measurement using Bruker's MultiMode 8 AFM. The tip used is PFQNE-AL, which has a frequency of 300 kHz and a spring constant of 0.8 N/m, and the tip radius is 5 nm. As it is commonly known, KPFM measures the work function of the sample which is an external potential that offsets the potential difference between the tip and the sample surface. Measurement methods are divided into AM-KPFM and FM-KPFM according to the feedback method. Among them, FM-KPFM has a better lateral resolution, making it easy to measure small \moire superlattices\cite{zerweck2005accuracy}. The \Moire potential can be measured by the difference in surface potential measured in AB and BA domains at hBN, and when positive bias is performed through the tip, the high potential area corresponds to the BA domain and the low potential area corresponds to the AB domain\cite{woods}\cite{chiodini2022moire}.\\

\textbf{PL measurements and Hyperspectral imaging}
    We used home-built confocal microscope set up with continuous-wave (CW) green laser (532 nm) for excitation. An objective lens with NA = 0.9 allowed us to reach laser spot size of 1$\textrm{ μm}^{2}$. We used 42.9 μW for the exciton and trion state excitation and 386 μW for the Rydberg states. The set up is coupled with a EM-CCD for the high efficiency measurements. The sample is placed in a closed cycle cryostat (Montana) for the low temperature measurements. For the hyperspectral imaging, we used XY galvometer system in a 4f configuration.\\

\vspace{12pt}
\section{Acknowledgements}

This research was supported by the National Research Foundation of Korea (NRF) grant funded by the Korea government (MSIT) (2021K1A3A1A32084700, 2021R1A2C2093155, 2021M3H4A1A03054856, 2022R1A4A3030766, 2022M3H4A1A04096396, RS-2023-00254055). This work was supported by a grant from Kyunghee University in 2019 (KHU-20192441). 
Work at City College of New York was funded by the AFOSR Global grant - AFOSR-FA2386-21-1-4087 (B.D.), NSF DMR-2130544 (S.Y.) and the NSF MRSEC PAQM (DMR-2011738). 
This work was supported by the Korean NRF through the Grants No. 2020R1A2C3009142 (F.L.), 2020R1A5A1016518 (J.J.) and 2021K1A3A1A32101983 (J.J.).
Work at CUNY ASRC was funded by the NSF DMR-2044281 (G.G.).
K.W. and T.T. acknowledge support from the JSPS KAKENHI (Grant Numbers 21H05233 and 23H02052) and World Premier International Research Center Initiative (WPI), MEXT, Japan. 

\pagebreak
% References
\medskip

% Use the following code if you wish to generate your ` with BibTeX;
% replace the string "MSP-template" below with the name(s) of
% the BibTeX data base(s) you want to use.
% The resulting bibliography-output (the content of the .bbl file)
% must be pasted back into this file before submission.
% Please also include your BibTeX data base file(s) in your submission
% so that we can re-run BibTeX if necessary.
%
%\bibliography{MSP-template}

\bibliography{references}

%%%
%\textbf{References}\\

%1	((Journal articles)) a) A. B. Author 1, C. D. Author 2, Adv. Mater. 2006, 18, 1; b) A. Author 1, B. %Author 2, Adv. Funct. Mater. 2006, 16, 1.\\
%2	((Work accepted)) A. B. Author 1, C. D. Author 2, Macromol. Rapid Commun., DOI: 10.1002/marc.DOI.\\
%3	((Books)) H. R. Allcock, Introduction to Materials Chemistry, Wiley, Hoboken, NJ, USA 2008.\\
%4	((Edited books or proceedings volumes)) J. W. Grate, G. C. Frye, in Sensors Update, Vol. 2 (Eds: H. %Baltes, W. Göpel, J. Hesse), Wiley-VCH, Weinheim, Germany 1996, Ch. 2.\\
%5	((Presentation at a conference, proceeding not published)) Author, presented at Abbrev. Conf. %Title, Location of Conference, Date of Conference ((Month, Year)).\\
%6	((Thesis)) Author, Degree Thesis, University (location if not obvious), Month, Year.\\
%7	((Patents)) a) A. B. Author 1, C. D. Author 2 (Company), Country Patent Number, Year; b) W. %Lehmann, H. Rinke (Bayer AG) Ger. 838217, 1952.\\
%8	((Website)) Author, Short description or title, URL, accessed: Month, Year.\\
%9	…((Please include all authors, and do not use “et al.”))\\

% Figures/tables and captions
% Permission statements are required for all figures reproduced or adapted from previously published articles/sources. Please also ensure that all necessary permissions to reproduce images have been received
% Please remove these statements for original figures

\pagebreak

\begin{figure}
  \includegraphics[width=\linewidth]{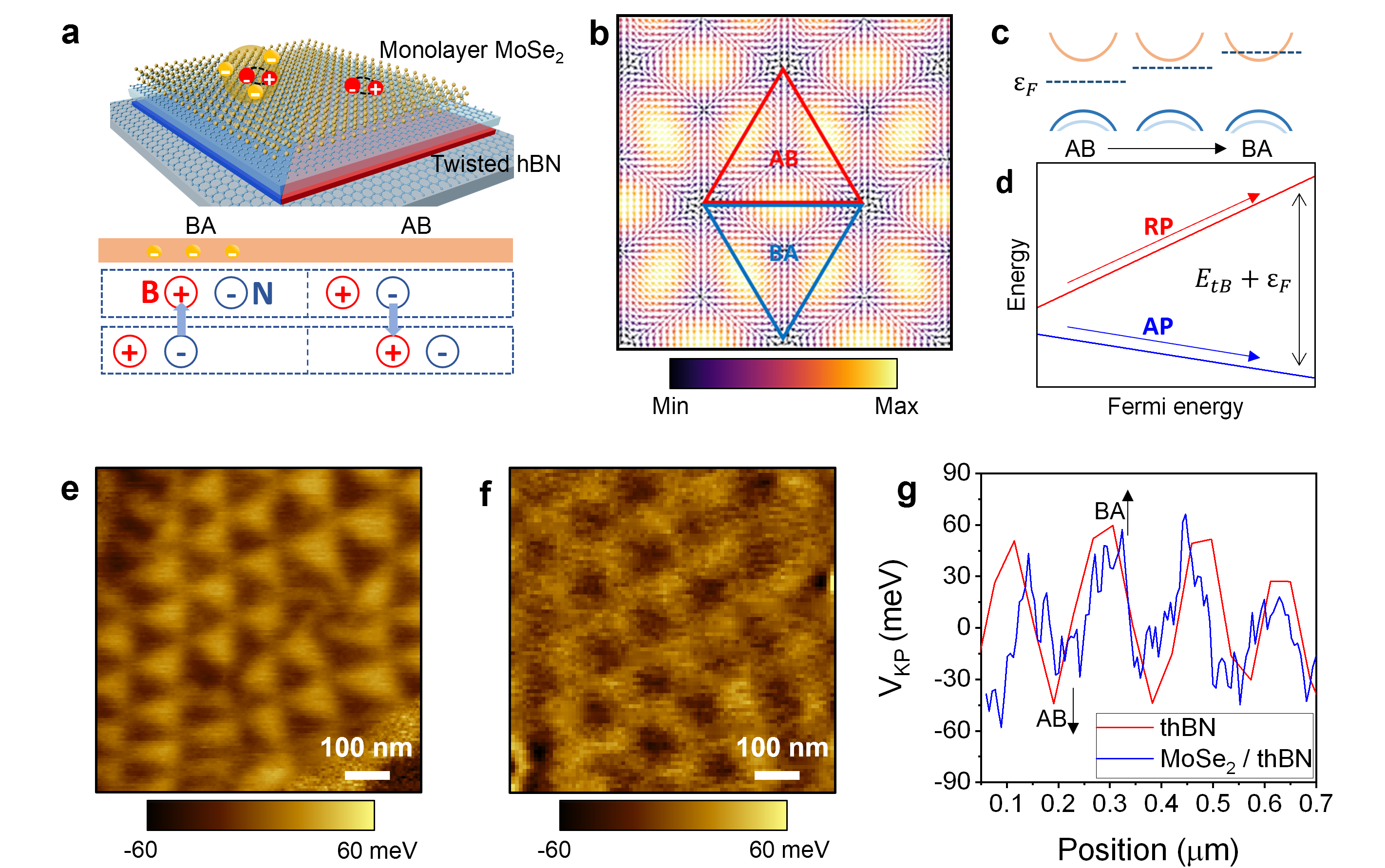}
  \caption{thBN \moire superlattice imprinting on the monolayer \MoSe. (a) Schematic of the sample. When two hBN layers are aligned in parallel, they form triangular antiferroelectric \moire domains consists AB and BA domain at the interface. In the schematic, the lower part depicts the atomic arrangement at the boundary between the two hBN layers. Due to the relaxed atomic arrangement of Boron and Nitrogen, a downward dipole moment exists in the AB domain, while an upward electric dipole exists in the BA domain. Due to the electric dipole in each domain, an electric field directed in one direction is formed at the domain boundary. On the surface, \MoSe placed on the thBN and electron is redistributed by electric fields. The excited Fermi sea in the BA domain coupled with excitons to form exciton polarons. (b) The in-plane electric dipole schematic of thBN. When \MoSe is positioned on top of thBN, the direction of the electric field is from the BA domain to the AB domain, leading to an increase in electron density and Fermi energy in the BA domain. On the other hands, when \MoSe is positioned below thBN, there is an increase in electron density and Fermi energy in the AB domain. (c) The schematic showing the Fermi energy variation depending on the position of \MoSe on thBN. (d) Schematic representation of repulsive polaron (RP) and attractive polaron (AP) energies depending on Fermi energy. (e) KPFM measurements of \moire superlattice on thBN. The superlattice size is 161±30 nm, corresponding to the twist angle around 0.089 degrees. (f) KPFM measurements of \moire superlattice on \MoSe/ thBN. (g) The change in \moire potential difference before and after \MoSe transfer in Figure 1(e), (f). While the potential difference remains relatively constant around 110 meV, it is notable that there is an increase in noise.}
  \label{fig:boat1}
\end{figure}

\begin{figure}
  \includegraphics[width=\linewidth]{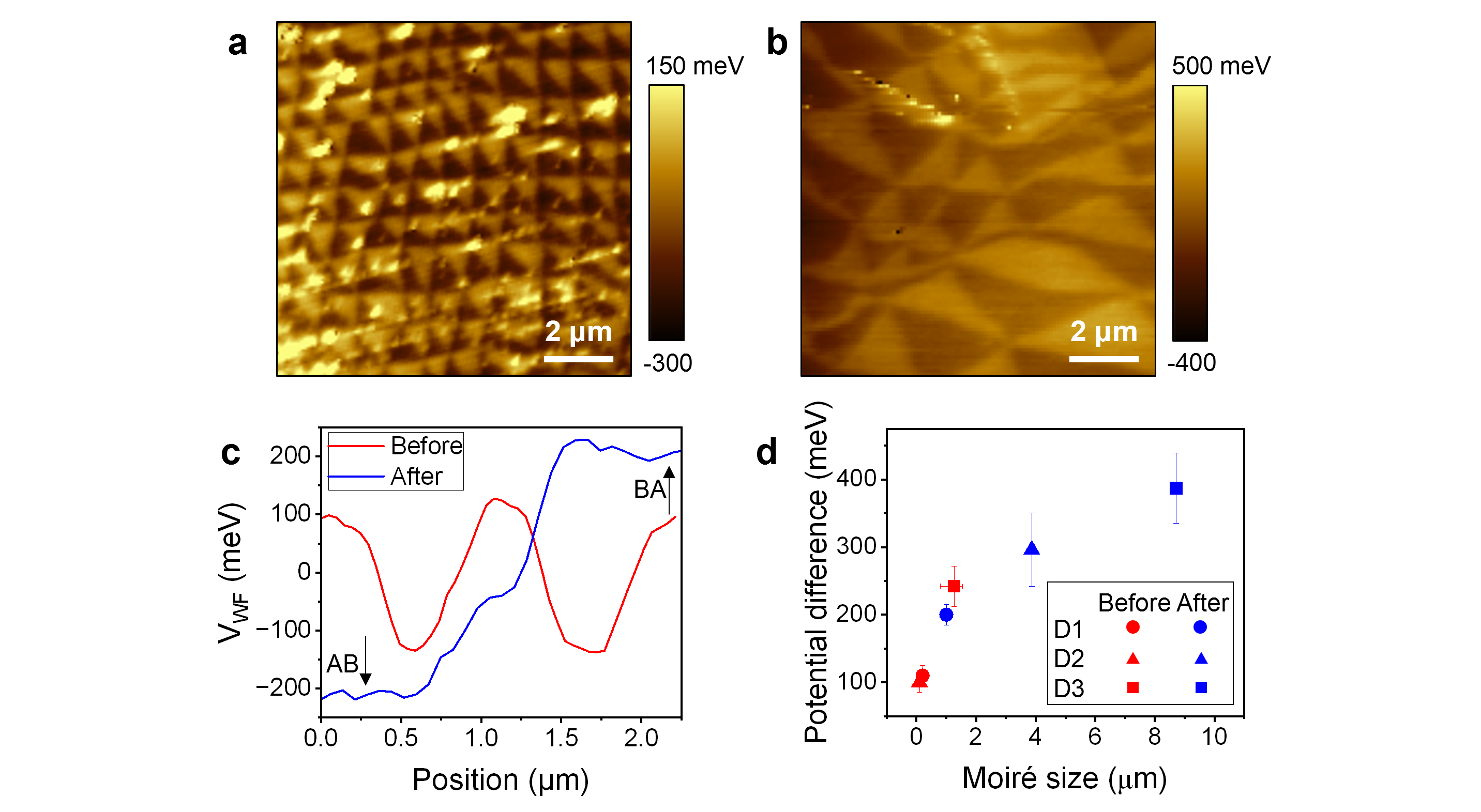}
  \caption{\Moire potential maximize after repeating the transfer process. (a) KPFM measured after the first transfer of thBN. The \moire size ranges from 793 nm to 1266 nm, it is the largest size among the samples we stacked, corresponding to angles between 0.011 and 0.019 degrees. (b) KPFM measured after the second transfer process on the same {SiO$_2$}/Si substrate. The \moire size increased significantly, ranging from approximately 1310 to 8720 nm, an increase of over 10 times. The angle in this region is below 0.01 degrees. (c) The change in \moire potential difference the first and the second transfer process in Figure 2(a), (b). After the first transfer process, the potential is 240±30 meV (red line), and after the second transfer process, it increases to 387±52 meV (blue line), along with the increase in moiré size. (d) Potential difference versus \moire size for device 1 (circle), device 2 (triangle), device 3 (rectangle). After the second transfer process, we marked maximum \moire size values among various sizes.}
  
  \label{fig:boat2}
\end{figure}

\begin{figure}
  \includegraphics[width=\linewidth]{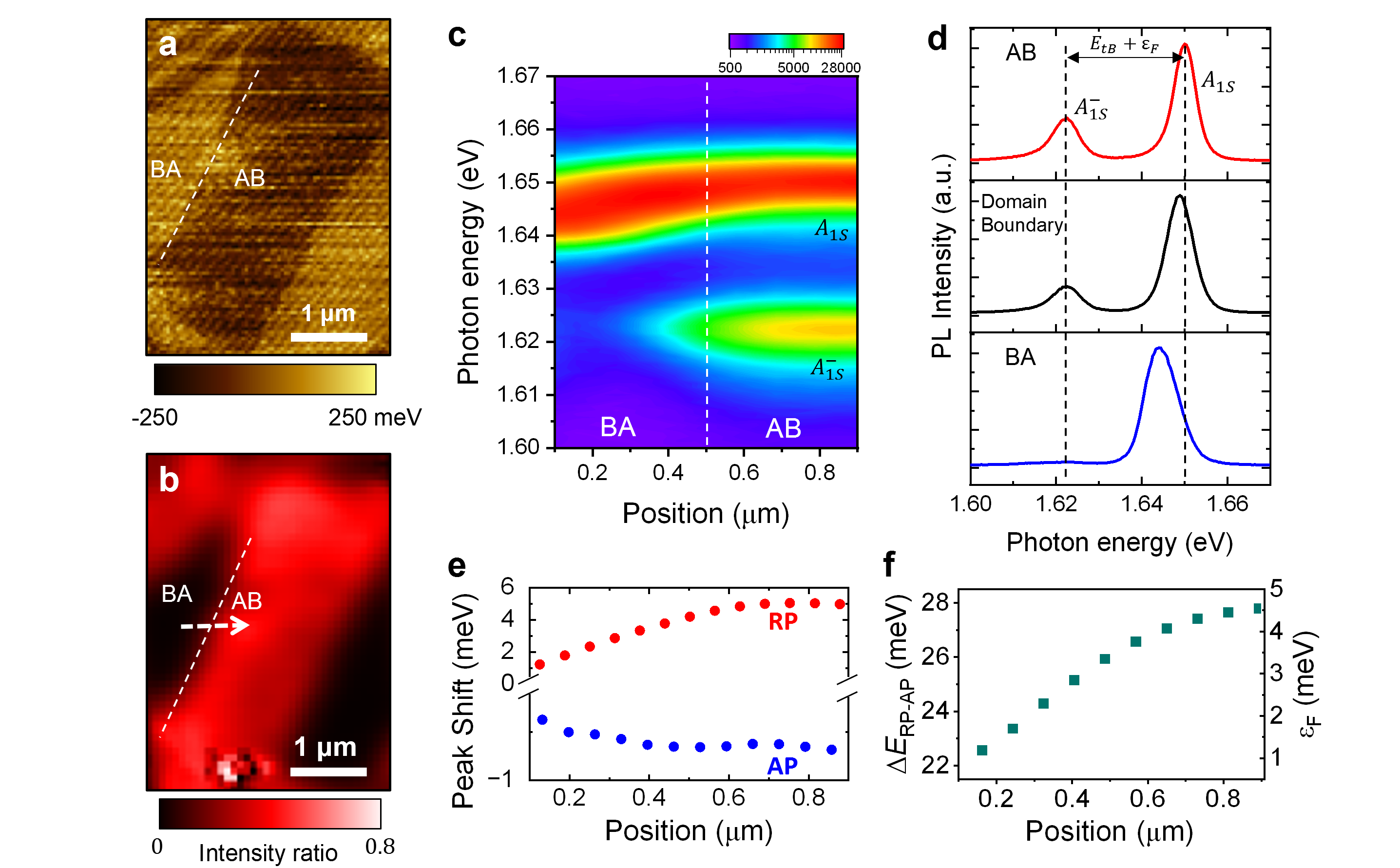}
  \caption{Hyperspectral imaging of exciton polaron in thBN/MoSe$_2$/hBN sample. (a) KPFM measurements of thBN/MoSe$_2$/hBN sample. (b) Hyperspectral PL imaging with excitation CW laser of 2.33 eV at the same location as Figure 3(a). Measured temperature is 8 K. Trion state ($A_{1S}^-$) intensity normalized by exciton state ($A_{1S}$) intensity. In the AB domain, a high intensity ratio is observed, while in the BA domain, a value close to 0 is shown, allowing the \moire pattern to be clearly seen in the PL. (c) Contour plot of PL spectra along the white arrow in Figure 3(b). (d) Normalized PL spectra of $A_{1S}^-$, $A_{1S}$ peak in AB, BA domain and near domain boundary. $A_{1S}^-$ position shifted from 1.623 eV to 1.622 eV from BA to AB domain. $A_{1S}$ position shifted from 1.644 eV to 1.650 eV from BA to AB domain. (e) Peak shift depending on the position. $A_{1S}^-$ shows red shift corresponding to attractive polaron. $A_{1S}$ shows blue shift correspoding to repulsive polaron. (f) Energy difference between attractive polaron and repulsive polaron. Fermi level corresponding to the energy difference.
}
  \label{fig:boat3}
\end{figure}

\begin{figure}
  \includegraphics[width=\linewidth]{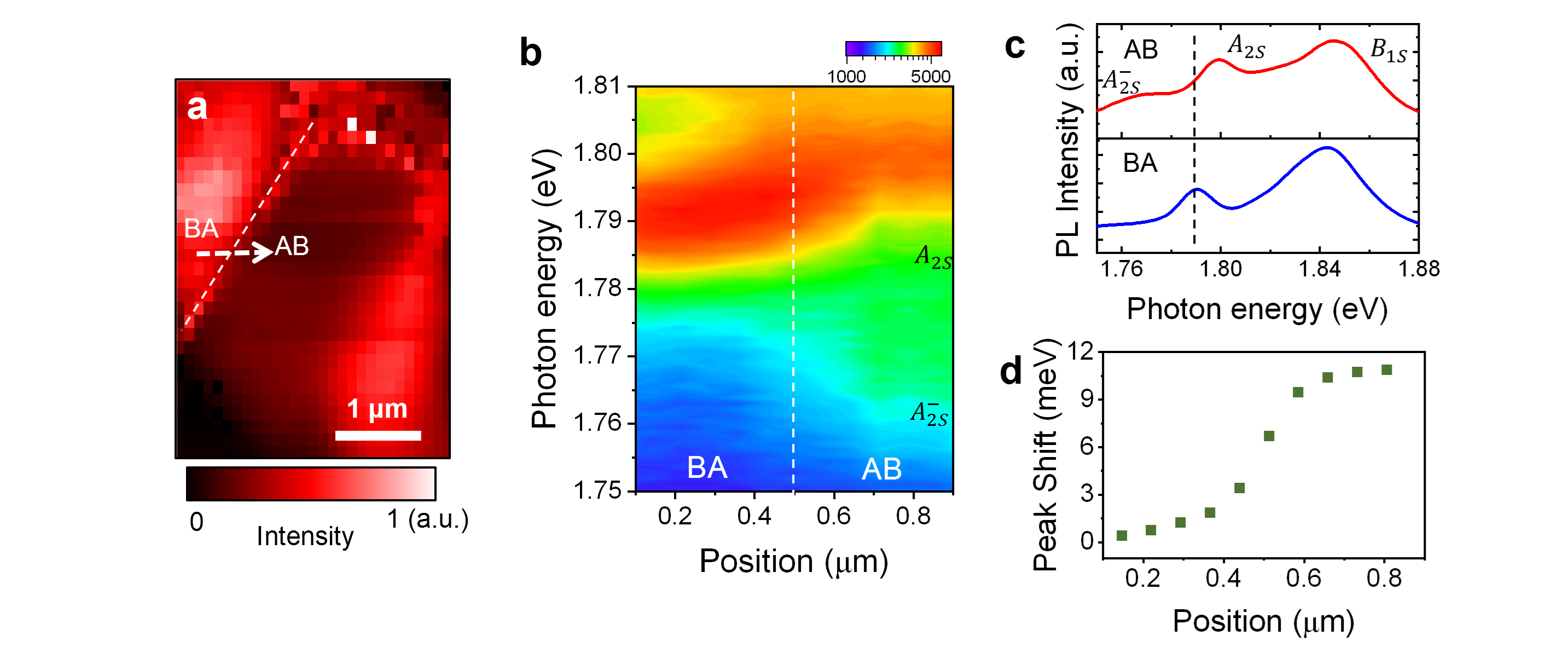}
  \caption{Rydberg exciton polaron states. (a) Hyperspectral PL imaging with excitation CW laser of 2.33 eV at the same location as Figure 3(a). Measured temperature is 8 K. $A_{2S}$ intensity is plotted. (b) Contour plot of PL spectra along the white arrow in Figure 4(a). (c) Normalized PL spectra of $A_{2S}^-$, $A_{2S}$, $B_{1s}$ in AB and BA domain. $A_{2S}$ position shifted from 1.790 eV to 1.799 eV from BA to AB domain. (d) Peak shift depending on the position. Excited exciton polaron states shows larger peak shift compared to exciton polaron.}
  \label{fig:boat4}
\end{figure}

\begin{figure}
  \includegraphics[width=\linewidth]{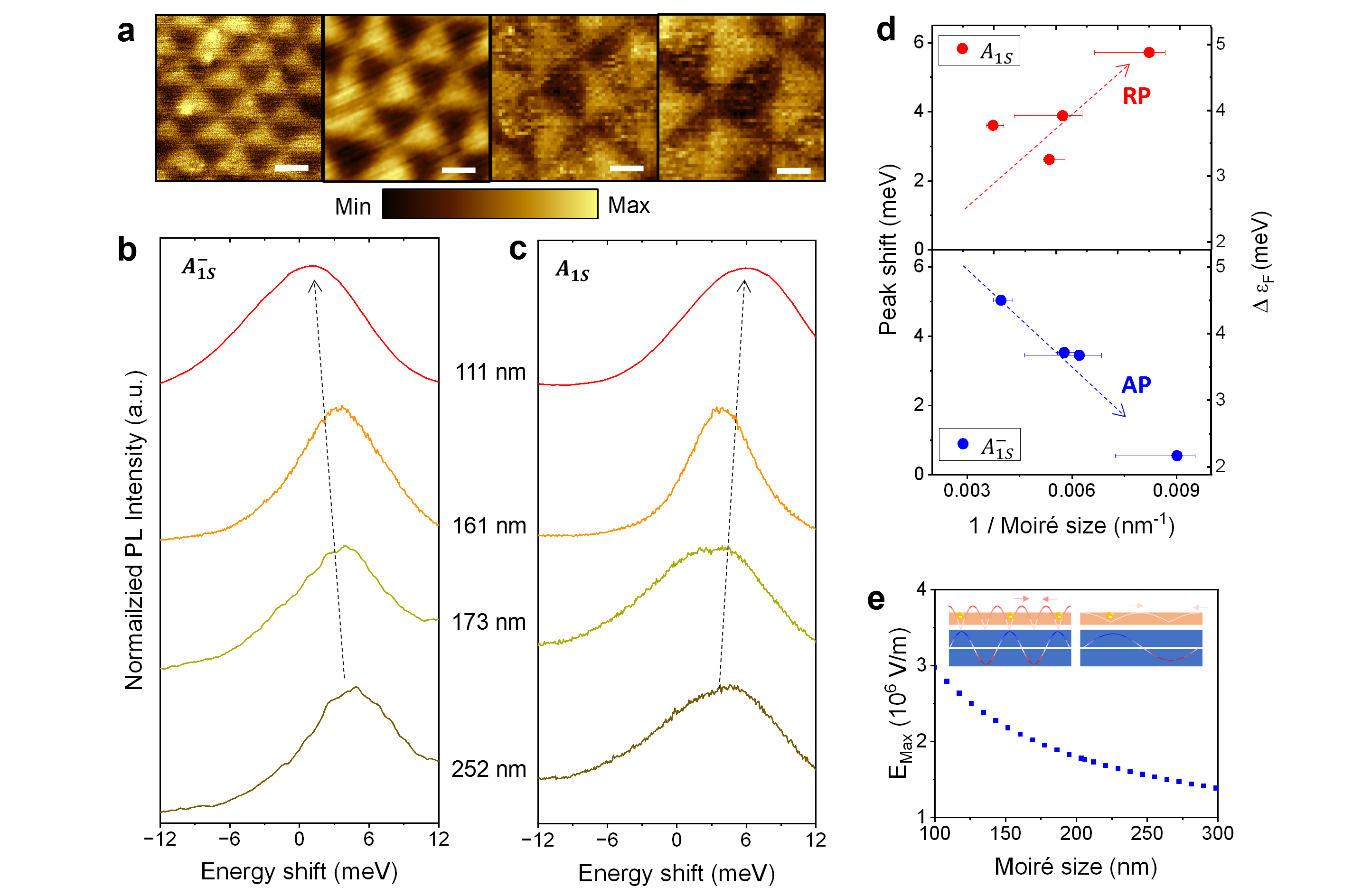}
  \caption{Exciton polaron peak shift induced by \moire superlattice. (a) KPFM on thBN, where the PL of \MoSe was measured. The \moire sizes, from left to right, are approximately 111 nm, 161 nm, 173 nm, and 252 nm. The scale bar is 100 nm. (b) Normalized $A_{1S}^-$ peak shift depending on the \moire size. (c) Normalized $A_{1S}$ peak shift depending on the \moire size. Excitation laser is 2.33 eV CW laser and the sample temperature is 8 K. We normalized the energy difference from the moiré-less region for each sample. (d) The PL spectra peak position in Figure 4(b) and (c) as a function of 1/\moire sizes. Repulsive polaron exhibits a consistent blue shift, while the attractive polaron displays a red shift. Fermi level change corresponding to the repulsive polaron peak shift. (e) The strength of electric field as a function of the \moire size, calculated from the potential differences shown in Figure 2(d). The inset figure indicates that a large electric field is applied at small \moire sizes, causing charge redistribution in specific domains of the MoSe$_2$, which increases the Fermi level.}
  \label{fig:boat5}
\end{figure}
%\begin{table}
% \caption{Table 1 caption}
%  \begin{tabular}[htbp]{@{}lll@{}}
%    \hline
%    Description 1 & Description 2 & Description 3 \\
%    \hline
%    Row 1, Col 1  & Row 1, Col 2  & Row 1, Col 3  \\
%    Row 2, Col 1  & Row 2, Col 2  & Row 2, Col 3  \\
%    \hline
%  \end{tabular}
%\end{table}

% Please provide Biographies and photos for Essays, Feature Articles, Progress Reports, Reviews, and Perspectives for those authors who should be highlighted  
% These should be at most 100 words long
% For other article types this section can be removed
% Photographs should be 40mm broad and 50 mm high

%\begin{figure}
%  \includegraphics{bio-placeholder.jpg}
%  \caption*{Biography}
%\end{figure}

%\begin{figure}
%  \includegraphics{bio-placeholder.jpg}
%  \caption*{Biography}
%\end{figure}

%\begin{figure}
%  \includegraphics{bio-placeholder.jpg}
%  \caption*{Biography}
%\end{figure}

%\begin{figure}
%  \includegraphics{bio-placeholder.jpg}
%  \caption*{Biography}
%\end{figure}

% Table of contents entry should be 50 - 60 words long
% Image should be 55 mm broad and 50 mm high or 110 mm broad and 20 mm high

%\begin{figure}
%\textbf{Table of Contents}\\
%\medskip
%  \includegraphics{toc-image.png}
%  \medskip
%  \caption*{ToC Entry}
%\end{figure}

\counterwithout{figure}{section} 
\renewcommand{\thefigure}{S\arabic{figure}}
\setcounter{figure}{0}  

\section{Supporting Information}\label{secA1}
\vspace{100pt}
\begin{figure}
  \includegraphics[width=\linewidth]{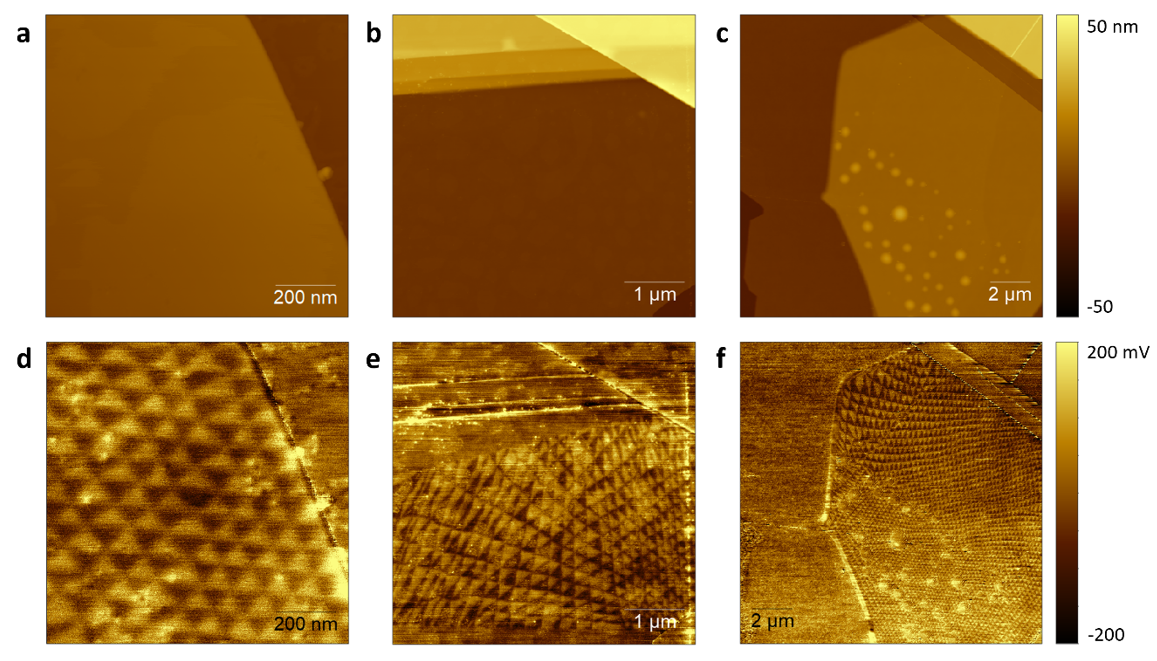}
  \caption{First transfer process of thBN. We stacked the hBN flake aligned nearly perfectly in their as-exfoliated state.  (a), (b), and (c) show the topography. The thickness of the top hBN 5 nm, 7 nm, 9 nm, and the bottom hBN is 5 nm, 14 nm, 16 nm. Panels (d), (e), and (f) represent KPFM measurements, revealing moiré sizes and stacking angles: (d) moiré size of 111 nm and stacking angle of 0.129 degrees; (e) moiré sizes ranging from 173 nm to 658 nm, with corresponding stacking angles ranging from 0.0218 to 0.0829 degrees; and (f) moiré sizes ranging from 260 nm to 370 nm, with corresponding stacking angles ranging from 0.0388 to 0.0552 degrees.}
  \label{fig:boat1}
\end{figure}

\pagebreak

\begin{figure}
  \includegraphics[width=\linewidth]{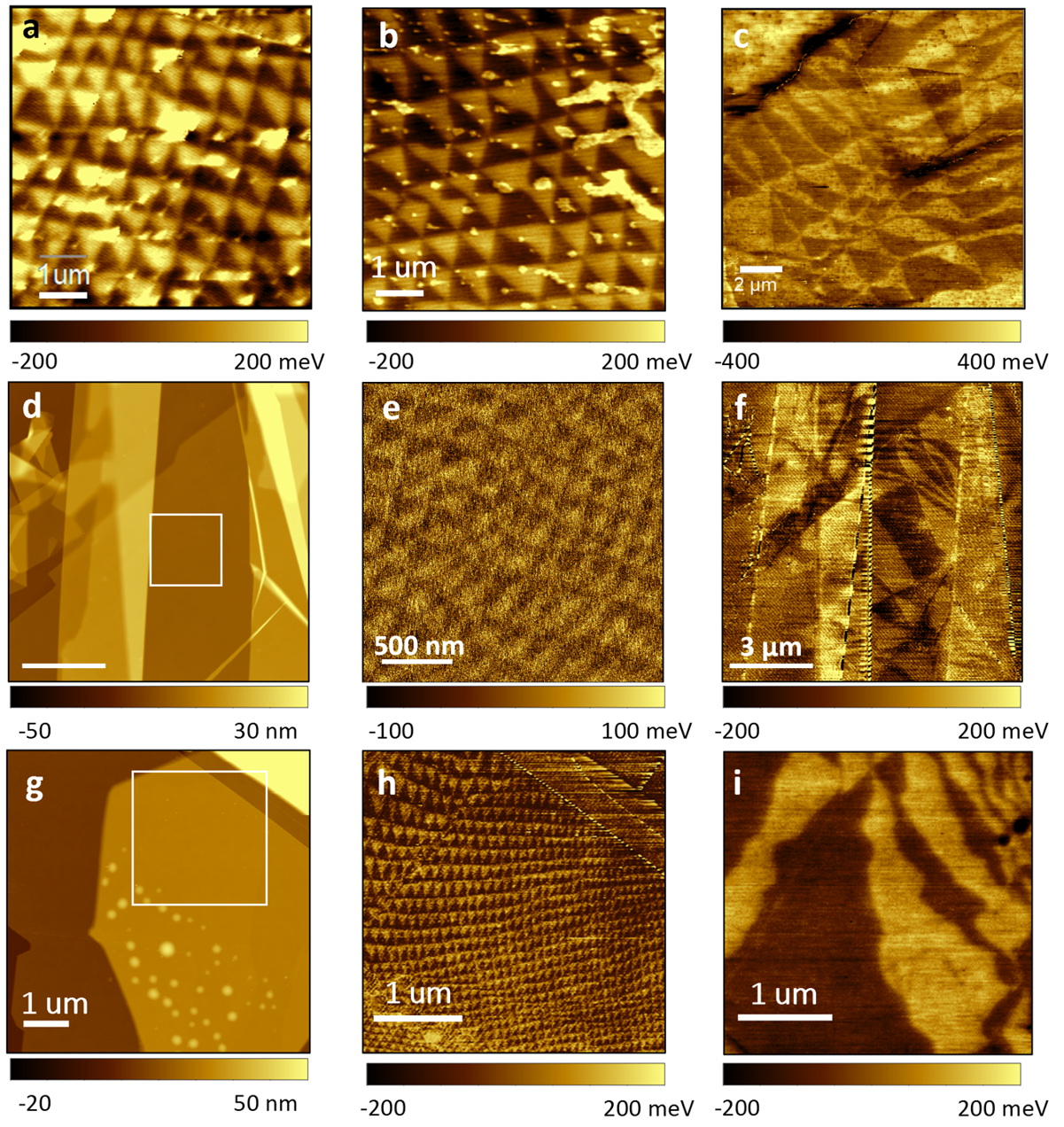}
  \caption{Second transfer process of thBN. (a) KPFM image immediately after stacking using the PC dry transfer method. (b) After the annealing at 200 °C, 1h with 100 sccm nitrogen flow. We confirmed that the moire size does not change during the annealing process. (c) Moiré size is larger after the second transfer process on the same SiO$_2$ / Si substrate. (d, g) Topography of the thBN. KPFM measured before (e, h)  and after (f, i) the second transfer process on white box at (d, g).}
  \label{fig:boat2}
\end{figure}

\begin{figure}
  \includegraphics[width=\linewidth]{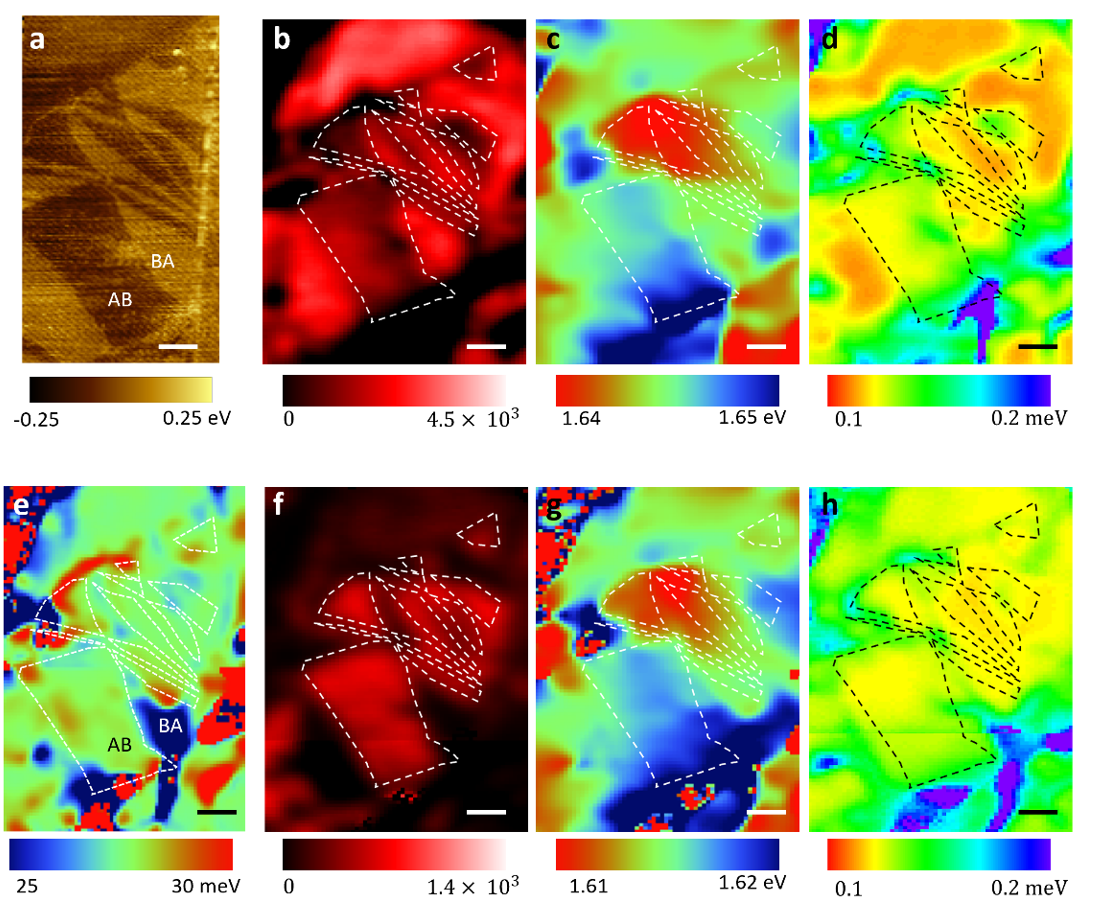}
  \caption{Hyperspectral of thBN / MoSe$_2$ / hBN.  (a) KPFM image of sample. Repulsive polaron (b) intensity, (c) energy, (d) linewidth. (e) The difference between repulsive and attractive polaron energy. Attractive polaron (f) intensity, (g) energy, (h) linewidth. Scale bar is 1 µm. }
  \label{fig:boat3}
\end{figure}

\begin{figure}
  \includegraphics[width=\linewidth]{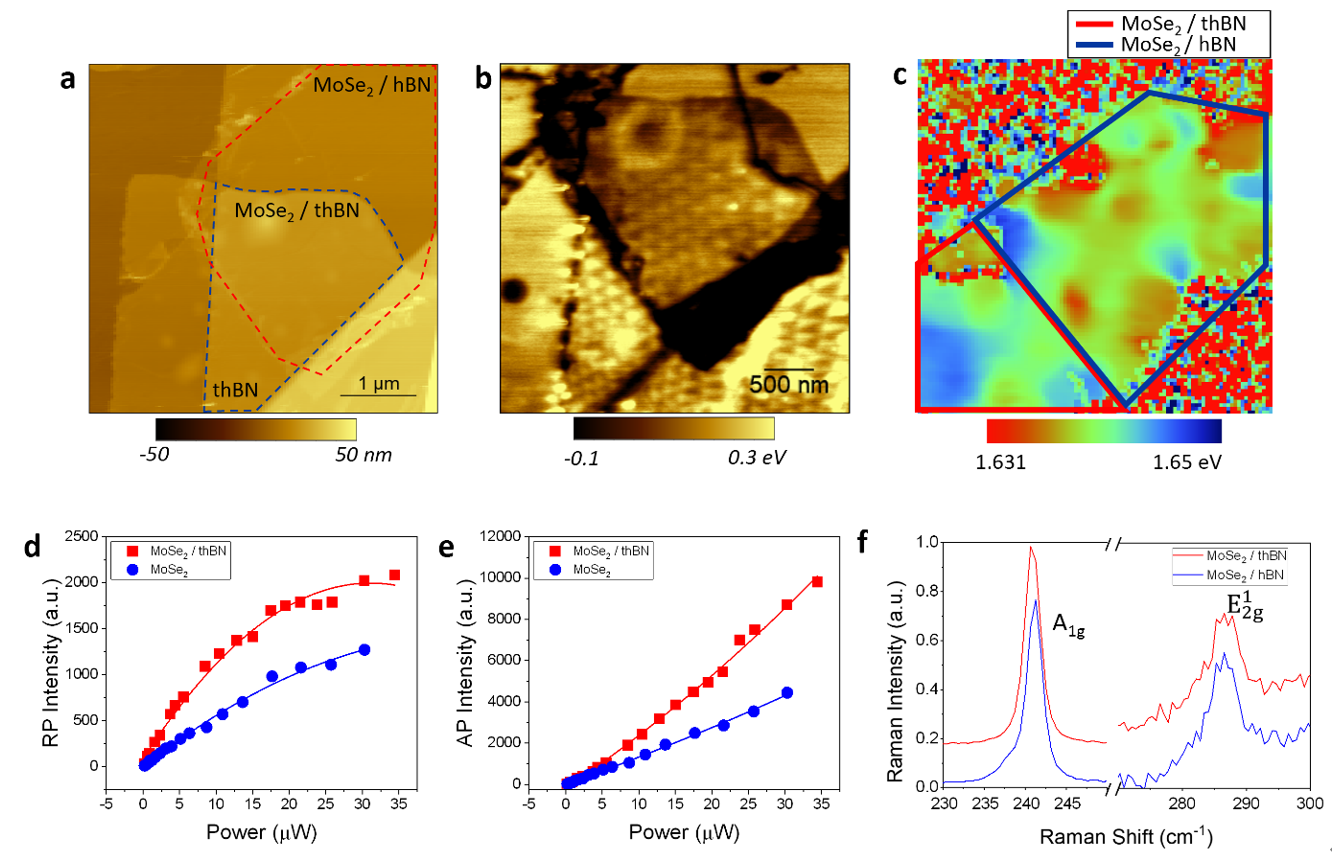}
  \caption{Hyperspectral of MoSe$_2$ / thBN.  (a) AFM topography image. MoSe$_2$ area is marked with a red dotted line. thBN area is marked with a blue dotted line. (b) KPFM image. The area above MoSe$_2$ has lower potential due to differences in work function, but the moiré pattern is identified the same. (c) Hyperspectra PL mapping image of repulsive polaron energy. Blue shift is observed in the thBN area. The power dependence of (d) repulsive polaron and (e) attractive polaron. The solid line is the value fitted by the parabola function. (f) Raman spectrum of MoSe$_2$ on thBN and hBN.}
  \label{fig:boat4}
\end{figure}

\end{document}